\pdfoutput=1
\documentclass[
  reprint,
  aps,
  prl,
  groupedaddress,
  amsmath,amssymb,
  floatfix
]{revtex4-2}
\usepackage{graphicx}
\usepackage{dcolumn}
\usepackage{bm}
\usepackage[dvipsnames]{xcolor}

\usepackage[T1]{fontenc}
\usepackage[utf8]{inputenc}
\usepackage{lmodern}

\newcommand{\dd}{\mathrm{d}}

\newcommand{\Tc}{T_{\mathrm{c}}}
\newcommand{\Tind}{T_{\mathrm{ind}}}
\newcommand{\Tdep}{T_{\mathrm{dep}}}
\usepackage[hidelinks]{hyperref}

\begin{document}

\title{Canonical Criterion for Third-Order Transitions}

\author{Fangfang Wang$^{1,5}$}

\author{Wei Liu$^{2,1}$}
\email{weiliu@xust.edu.cn}
\thanks{Corresponding author.}

\author{Kai Qi$^{4}$}

\author{Zidong Cui$^{3}$}

\author{Ying Tang$^{3}$}
\email{jamestang23@gmail.com}
\thanks{Corresponding author.}

\author{Zengru Di$^{1,5}$}

\affiliation{%
$^1$Department of Systems Science, Faculty of Arts and Sciences, Beijing Normal University, Zhuhai 519087, China \\
$^2$College of Science, Xi'an University of Science and Technology, Xi'an 710600, China \\
$^3$Institute of Fundamental and Frontier Sciences, University of Electronic Science and Technology of China, Chengdu 611731, China \\
$^4$2020 X-Lab, Shanghai Institute of Microsystems and Information Technology, Chinese Academy of Sciences, Shanghai 200050, China \\
$^5$School of Systems Science, Beijing Normal University, Beijing 100875, China
}

\begin{abstract}
Microcanonical inflection-point analysis (MIPA) identifies third-order transitions from derivatives of the microcanonical entropy, but whether such transitions admit a direct canonical formulation has remained unclear. Here we establish a fluctuation-based canonical framework for third-order transitions through a cumulant-ratio criterion whose signed extrema define their canonical counterparts and, in the single-saddle regime, are asymptotically linked to microcanonical classification. Because the criterion depends only on energy cumulants, it avoids explicit density-of-states reconstruction and remains operational in nonequilibrium steady states. Physically, it reveals dependent and independent third-order transitions as fluctuation reorganizations around low-order transitions, namely disordered-side precursors and ordered-side restructuring. Benchmarks on Onsager’s two-dimensional Ising solution, finite size Potts models, and a driven nonreciprocal Ising model show that the framework is theoretically grounded and broadly applicable.
\end{abstract}

\maketitle
Phase transitions are traditionally characterized by singularities in thermodynamic potentials or response functions in the thermodynamic limit~\cite{Landau1999PhaseTransitionsCiSE,Wilson1983,Ehrenfest1933Amsterdam}. 
However, in finite systems these singularities are rounded~\cite{Fisher1972FiniteSizeScalingPRL}, and the physics near a major transition can include additional transitions without singularity. 
In particular, finite systems often exhibit higher-order features, such as precursor-like fluctuations on the disordered side and partial ordered-side reorganization~\cite{BelHadjAissa2020MicrocanonicalDerivatives,Sitarachu2020ThirdOrderIsing,Sitarachu2022IsingThirdOrder,Liu2025PottsGeometry}. 
These higher-order transitions encode meaningful information about cooperative rearrangements, defect formation, and mesoscopic restructuring that are invisible to conventional low-order measures~\cite{Qi2019FlexiblePolymers,Aierken2023SemiflexiblePolymers}. 
Understanding these features is essential for capturing the full complexity of finite system collective behavior and for connecting microscopic fluctuations to mesoscopic organization.

A key question is whether higher-order transitions can be formulated in canonical terms, where fluctuations are naturally defined, experimentally or numerically accessible, and directly linked to mesoscopic reorganization. MIPA addresses this via the curvature of the microcanonical entropy, identifying dependent third-order transitions above and independent ones below a lower-order transition, corresponding to precursor-like and ordered-side restructuring~\cite{Gross2001MicrocanonicalThermodynamics,Qi2018MIPAClassification}. While applied to Potts and Baxter--Wu models, XY lattices, compact electrodynamics, and protein-folding systems~\cite{liu2022pseudo,wangff,DiCairano2024Topological,DiCairano2022Topological,BelHadjAissa2021KT_XY}, MIPA remains intrinsically microcanonical. 
Its implementation often requires explicit density-of-states (DOS) reconstruction, which is computationally costly and often unavailable in nonequilibrium steady states~\cite{Bachmann2025Histogram,Wang2001MultipleRangeRW,Wang2001FlatHistogramPRE,Berg1991MulticanonicalAlgorithms,Berg1992MulticanonicalEnsemble}.
Even at equilibrium~\cite{Fruchart2021NonreciprocalNature,GarciaLorenzana2025NonreciprocalSpinGlass,Loos2020IrreversibilityNJP,Avni2025NonreciprocalIsing}, the connection between third-order transitions and mesoscopic fluctuation reorganizations around low-order transitions is not explicit in canonical language.

These limitations motivate developing a canonical approach, where higher-order transitions can be directly extracted from energy fluctuations without reconstructing the DOS. 
In the following, we establish a fluctuation-based canonical framework for third-order transitions using the energy-cumulant $\Xi(T)={\kappa_3(T)}/{[\kappa_2(T)]^3}$, where $\kappa_n(T)$ is the $n$th cumulant of total energy.
Signed local extrema of $\Xi(T)$ define independent and dependent third-order transitions, and in the single-saddle regime they are asymptotically linked to microcanonical classification. 
In equilibrium, $\Xi(T)$ can be obtained directly from canonical fluctuations or, when $g(E)$ is available, by standard reweighting. 
This framework requires only canonical cumulants, remains operational in nonequilibrium steady states, and naturally makes dependent and independent third-order transitions explicit as disordered-side precursors and ordered-side fluctuation reorganizations around conventional low-order transitions.

\begin{figure*}[t]
  \centering  \includegraphics[width=\textwidth,height=0.32\textheight,keepaspectratio]{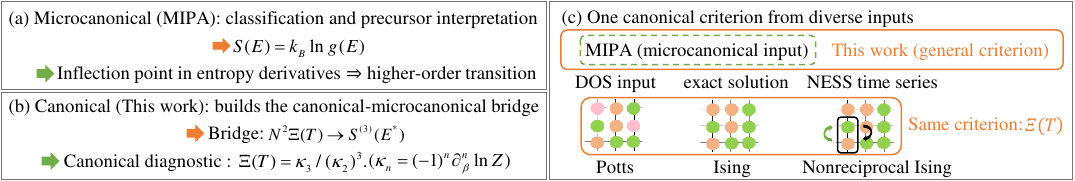}
\caption{Roadmap for the energy-cumulant diagnostic and its relation to MIPA. 
(a) MIPA classifies higher-order transitions via derivatives of $S(E)=k_B\ln g(E)$, where $g(E)$ is the DOS, identifying the dependent transition as a precursor and the independent one as an ordered-side reorganization. 
(b) $\Xi(T)$ establishes $N^2\Xi(T)\to s^{(3)}(e^*)$ in the single-saddle regime, linking signed extrema to fluctuations and providing a canonical-theory basis for third-order transitions. 
(c) Unlike MIPA, $\Xi(T)$ applies to both canonical and microcanonical ensembles. Benchmarks: Ising model, Potts models and nonreciprocal Ising model.}
  \label{fig:roadmap}
\end{figure*}
We benchmark $\Xi(T)$ in three representative settings. Onsager's exact two-dimensional Ising solution provides a sampling-free equilibrium reference and a clean, fully controlled test of the connection to microcanonical classification. Finite size Potts models spanning continuous and first-order regimes probe the robustness and finite size sensitivity of the criterion under rounding and ``phase'' coexistence. A driven nonreciprocal two-dimensional Ising model demonstrates extraction from nonequilibrium steady-state time series, where microcanonical input is unavailable, and reveals a precursor-like onset of cooperative dynamics and emergent synchronized behavior. Geometry-sensitive observables are used only as complementary probes of model-dependent mesoscopic restructuring; throughout this work, the primary diagnostic remains the cumulant-ratio criterion $\Xi(T)$.

\emph{Canonical Framework---} In the canonical ensemble, the partition function is
\begin{equation}
Z(\beta) = \sum_E g(E) e^{-\beta E},
\label{eq:Z}
\end{equation}
where $K(\beta)=\ln Z(\beta)$ is the cumulant-generating function. Here, $E$ is the total energy and $\beta \equiv 1/(k_B T)$ is the canonical inverse temperature, with $k_B$ set to unity. We distinguish $\beta$ from the microcanonical inverse temperature $\beta_\mu(E)\equiv {\dd S}/{\dd E}$, where $S(E)=k_B\ln g(E)$ is the microcanonical entropy. The energy cumulants are
$\kappa_n(\beta)=(-1)^n \partial_\beta^n K(\beta)$, so that $U\equiv \langle E\rangle=-K'(\beta)$, $\kappa_2=K''(\beta)$, and $\kappa_3=-K^{(3)}(\beta)$.

\begin{table*}[t]
  \centering
  \caption{Microcanonical criteria for third-order transitions and their canonical counterparts.}
  \label{tab:signals}
  \begin{tabular}{ccc}
    \hline
    Category & Dependent third-order & Independent third-order \\
    \hline
    Microcanonical (MIPA)~\cite{Qi2018MIPAClassification}
    & negative local maximum of $\dd^{3}S(E)/\dd E^{3}$
    & positive local minimum of $\dd^{3}S(E)/\dd E^{3}$  \\
    Canonical (This work)
    & negative local maximum of $\Xi(T)$
    & positive local minimum of $\Xi(T)$ \\
    \hline
  \end{tabular}
\end{table*}

We rewrite the partition sum as
\begin{equation}
Z(\beta) = \sum_E g(E) e^{-\beta E} \simeq \int \dd e\, \exp\!\big[N \phi_\beta(e)\big],
\label{eq:laplace}
\end{equation}
where $e\equiv E/N$ is the energy density, $N$ is the number of degrees of freedom, $\phi_\beta(e)\equiv s(e)-\beta e$, and $s(e)\equiv S(E)/N$. The saddle point $e^*$ satisfies $s'(e^*)=\beta$, where $s^{(n)}(e)\equiv {\dd^n s}/{\dd e^n}$. In the single-saddle regime, Laplace's method gives (Supplemental Material~\cite{SM}) $\kappa_2 \simeq -{N}/{s''(e^*)}$ and $\kappa_3 \simeq -{N\,s^{(3)}(e^*)}/{\big[s''(e^*)\big]^3}$. It follows that
\begin{equation}
N^2 \frac{\kappa_3}{\big[\kappa_2\big]^3} = s^{(3)}(e^*) + \mathcal O(1/N),
\label{eq:connection}
\end{equation}
which motivates the canonical cumulant ratio $\Xi(T)$ introduced here in Eq.~\eqref{eq:ourmethod}.

This relation also provides a practical equilibrium route: when an estimate of $g(E)$ is available, canonical cumulants and hence $\Xi(T)$ can be obtained by standard reweighting, without explicit canonical sampling. At fixed temperature $T$, the energy is distributed according to $P(E)={g(E)e^{-\beta E}}/{Z(\beta)}$. We therefore define
\begin{equation}
\Xi(T)=\frac{\kappa_3(T)}{\big[\kappa_2(T)\big]^3},
\label{eq:ourmethod}
\end{equation}
which probes the asymmetry of $P(E)$ through the third cumulant. The cubic normalization is chosen not to form the standardized skewness $\kappa_3/(\kappa_2)^{3/2}$, but to recover the single-saddle correspondence $N^2\Xi(T)\to s^{(3)}(e^*)$. Operationally, signed local extrema of $\Xi(T)$ identify third-order transitions: a negative local maximum identifies a dependent transition, whereas a positive minimum marks an independent one. Physically, these extrema mark turning points in the asymmetry of the energy distribution and hence in the associated fluctuation asymmetry.

Table~\ref{tab:signals} summarizes the microcanonical MIPA criteria and their canonical counterparts; a brief review of MIPA is given in Supplemental Material~\cite{SM}. Equation~\eqref{eq:connection} establishes the asymptotic bridge in the single-saddle regime. Beyond that regime, including finite size ``phase'' coexistence and nonequilibrium steady states, we treat signed extrema of $\Xi$ as a benchmarked operational diagnostic against independent references. Following Ref.~\cite{Qi2018MIPAClassification}, a dependent third-order transition can accompany a lower-order independent transition and typically appears on the higher-energy, higher-temperature side as a precursor. In this way, $\Xi$ makes dependent and independent third-order transitions explicit as precursor-like and ordered-side fluctuation reorganizations around conventional low-order transitions. The construction is in principle extensible to higher orders, but here we develop and benchmark only the third-order case.

\emph{Examples---} We benchmark $\Xi(T)$ in three settings. \emph{Onsager Ising Benchmark---} We first explicitly test the canonical diagnostic $\Xi(T)$ against Onsager's exact solution of the two-dimensional Ising model in the thermodynamic limit on a square lattice, as a well-established equilibrium reference. The Hamiltonian is~\cite{ising}
\begin{equation}
E=-J\sum_{\langle ij\rangle}s_i s_j,\qquad J>0,
\label{eq:ising}
\end{equation}
where $\langle ij\rangle$ denotes nearest-neighbor bonds counted once, and we set $J=1$. At zero field, the energy cumulants in the thermodynamic limit are obtained from Onsager's free-energy expression~\cite{Onsager1944}; numerical details are given in Supplemental Material~\cite{SM}. This benchmark provides a sampling-free reference for assessing the criterion and its correspondence with microcanonical classification.

\begin{figure}[t]
  \raggedright
  \includegraphics[width=1\columnwidth]{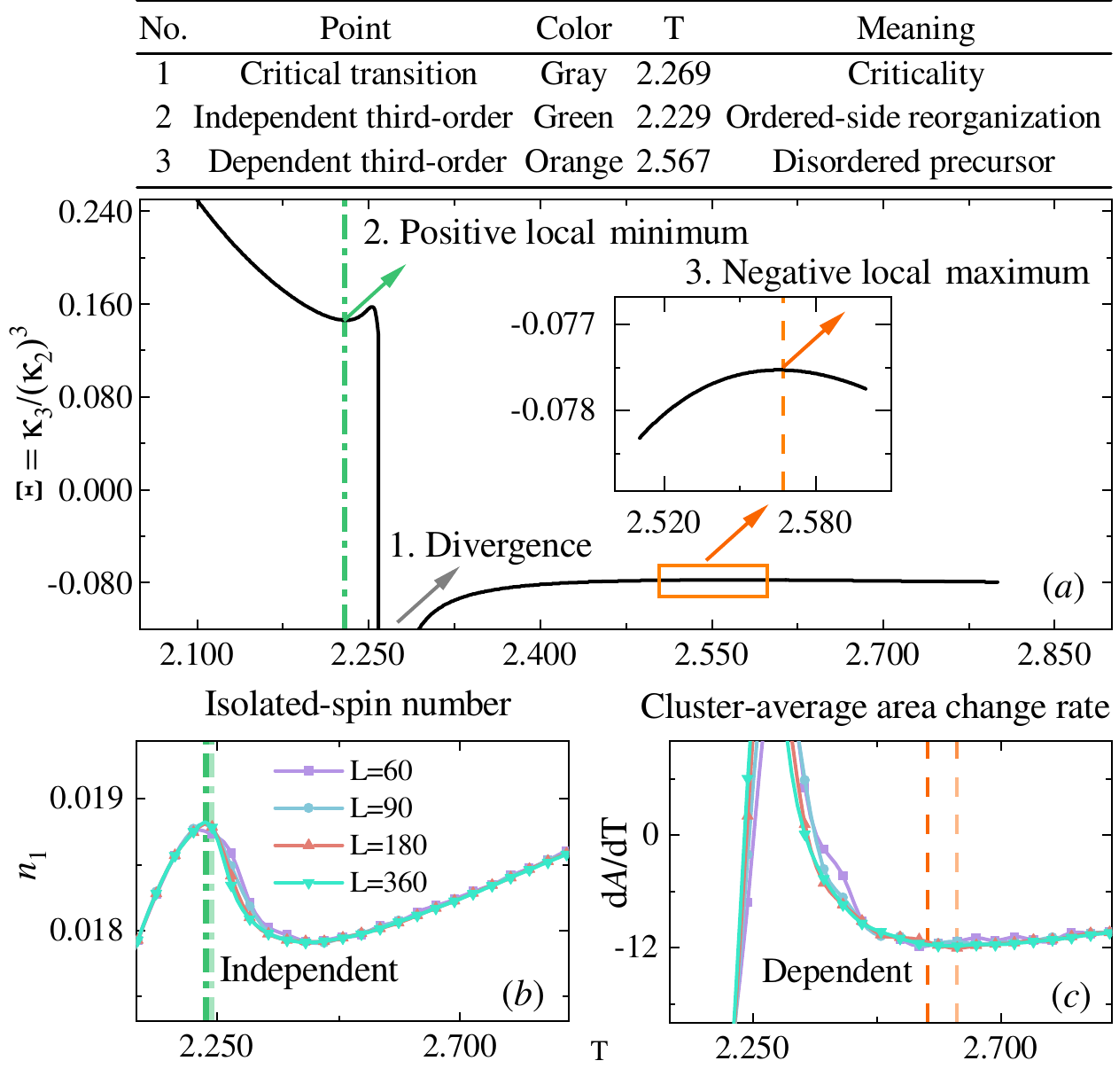}
\caption{Canonical third-order transitions and geometric corroboration in the 2D Ising model.
(a) Thermodynamic-limit $\Xi(T)$ from Onsager's exact free energy. The divergence at $T_c=2.269$ marks the critical transition; the positive minimum at $T_{\rm ind}=2.229$ (independent third-order, ordered-side reorganization) and negative maximum at $T_{\rm dep}=2.567$ (dependent third-order, disordered-side precursor) are exact and fully resolved.
(b,c) Complementary geometry-sensitive observables further corroborate panel (a). (b) Isolated-spin number $n_1$ shows a clear feature near $T_{\rm ind}$, consistent with the independent transition. (c) Cluster-averaged area change rate $\mathrm{d}A/\mathrm{d}T$ shows fastest variation near $T_{\rm dep}$, consistent with the dependent transition. Vertical lines mark $T_{\rm ind}$ (green) and $T_{\rm dep}$ (orange); dark lines indicate exact thermodynamic-limit values, lighter lines finite size estimates.}
  \label{fig:ising}
\end{figure}

As shown in Fig.~\ref{fig:ising}(a), $\Xi(T)$ exhibits two signed extrema on opposite sides of the critical divergence at $T_c$: a positive local minimum at $T_{\rm ind}=2.229$ and a negative local maximum at $T_{\rm dep}=2.567$, corresponding to independent (ordered-side) and dependent (disordered-side precursor) third-order transitions in this system. These extrema reproduce the microcanonical benchmark temperatures reported in the Supplemental Material~\cite{SM}. Because $\Xi(T)$ is derived from Onsager's exact free energy, the extrema reflect genuine canonical fluctuation structure and asymmetry, demonstrating that these third-order transitions have real physical significance and represent robust cooperative reorganization, rather than arising from finite size or numerical artifacts.

Figure~\ref{fig:ising}(b,c) provides direct complementary mesoscopic corroboration. Geometry-sensitive observables map intuitively to the canonical cumulant-ratio diagnostic: in finite size Ising lattice simulations, the peak of the isolated-spin number $n_1$ occurs at $T\approx 2.241$, while the corresponding local extremum of $\mathrm{d}A/\mathrm{d}T$ occurs at $T\approx 2.620$. Panel~(b) shows $n_1$, probing defect seeding and ordered-side restructuring, and panel~(c) shows $\mathrm{d}A/\mathrm{d}T$, which is more sensitive to clustering on the less-ordered side. Both observables serve as visual aids and corroborate the identification of the third-order transitions via $\Xi(T)$ without defining independent criteria.

\emph{Finite Size Potts---} To test finite size robustness, we study the ferromagnetic $q$-state Potts model on an $L\times L$ square lattice with periodic boundaries. In equilibrium, $\Xi(T)$ can be evaluated either directly from canonical energy-fluctuation data or by DOS-based reweighting when a reliable estimate of $g(E)$ is available (Supplemental Material~\cite{SM}). The Hamiltonian is~\cite{Potts1952,WuRMP1982}
\begin{equation}
E=-J\sum_{\langle ij\rangle}\delta(\sigma_i,\sigma_j),\qquad
\sigma_i\in\{0,1,\dots,q-1\},
\label{eq:potts_ham}
\end{equation}
where $\langle ij\rangle$ counts each nearest-neighbor bond once ($N_b=2L^2$). On the square lattice, the critical point obeys $\beta_c J=\ln(1+\sqrt{q})$, i.e., $T_c/J=1/\ln(1+\sqrt{q})$.

\begin{figure}[t]
  \centering
  \includegraphics[width=1\columnwidth]{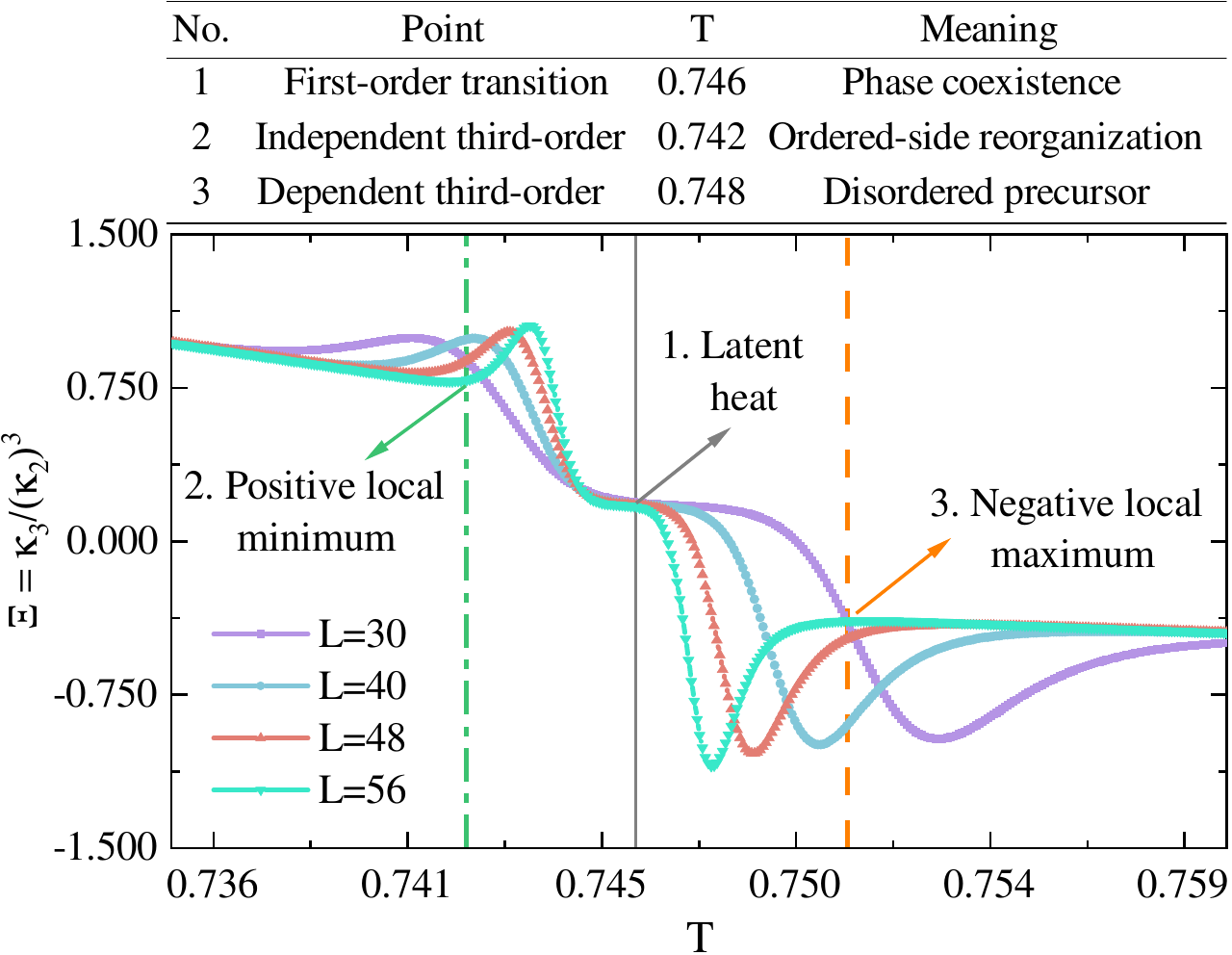}
  \caption{Finite size robustness of the canonical diagnostic $\Xi(T)$ in the square-lattice $q=8$ Potts model. From reweighting of a replica Wang--Landau DOS estimate, $\Xi(T)$ retains robust signed extrema around the first-order transition despite finite size and ``phase'' coexistence. The negative peak at $\Tdep(L)$ on the high-$T$ side identifies a precursor-like dependent third-order transition, whereas the positive dip at $\Tind(L)$ on the low-$T$ side identifies an independent ordered-side reorganization. Vertical dashed lines mark the $L=56$ values listed in the figure. Solid lines indicate the exact transition temperature $\Tc/J = 1/\ln(1+\sqrt{q})$.}
  \label{fig:potts}
\end{figure}

In the ``phase'' coexistence regime, where the single-saddle bridge no longer applies directly, the signed extrema of $\Xi(T)$ serve as a finite size canonical diagnostic. The square-lattice Potts transition is continuous for $q\le 4$ and first order for $q>4$ ($q=4$ marginal); here we focus on the first-order case $q=8$, with results for $q=3$, $4$, and $6$ in the Supplemental Material~\cite{SM}. Figure~\ref{fig:potts} shows that, despite finite size and ``phase'' coexistence, $\Xi(T)$ retains robust signed extrema, with a positive dip at low temperature identifying an independent ordered-side reorganization and a negative peak at high temperature identifying a precursor-like dependent transition. Crossings of $\Xi(T)$ for different $L$ yield a sharp estimate of $T_c$, consistent with the exact result $T_c/J = 1/\ln(1+\sqrt{q})$, and the corresponding microcanonical features match those in the Supplemental Material~\cite{SM}.

The third-order transitions in the $q = 8$ Potts model are basically consistent with MIPA results, confirming their physical significance. Similar transitions are observed in many other finite size lattice models~\cite{Bachmann2025Histogram,Sitarachu2020ThirdOrderIsing,liu2022pseudo,wangff}, and Onsager's exact solution shows that these transitions persist in the thermodynamic limit of the 2D Ising model~\cite{Sitarachu2022IsingThirdOrder}, indicating they are robust and worthy of further study. These results confirm that $\Xi(T)$ provides a stable canonical identification of third-order transitions across finite size systems and different model classes.
No uniform geometric proxy is imposed, as the most informative geometry-sensitive observable is model dependent. The exact $\kappa_3$--$C_V$ identity (Supplemental Material~\cite{SM}) further clarifies the thermodynamic content of the diagnostic. Overall, the cumulant-ratio criterion remains robust under finite size and ``phase'' coexistence, extending the canonical framework beyond the single-saddle regime.

\emph{Driven Nonreciprocal Ising---} 
Nonreciprocal interactions break detailed balance and can produce genuinely nonequilibrium phases and transition-like dynamical features~\cite{Fruchart2021NonreciprocalNature}. 
We consider a driven 2D nonreciprocal Ising model in which two spins $\sigma_i^A,\sigma_i^B \in \{\pm1\}$ occupy each lattice site and flip stochastically with Glauber-type rates~\cite{Glauber} determined by a local selfish energy,
\begin{equation}
E_i^\alpha = -J \sum_{j\in \mathrm{nn}(i)} \sigma_i^\alpha \sigma_j^\alpha
- K_m\,\varepsilon_{\alpha\beta}\,\sigma_i^\alpha \sigma_i^\beta,
\end{equation}
with $\alpha,\beta\in\{A,B\}$, $J,K_m>0$, and antisymmetric $\varepsilon_{\alpha\beta}$. Because the two species minimize different local energies, the dynamics explicitly violates detailed balance and relaxes to a nonequilibrium steady state (NESS)~\cite{Avni2025NonreciprocalIsing}, where the DOS is not defined and MIPA is not applicable.

\begin{figure}[t]
  \centering
  \includegraphics[width=1\columnwidth]{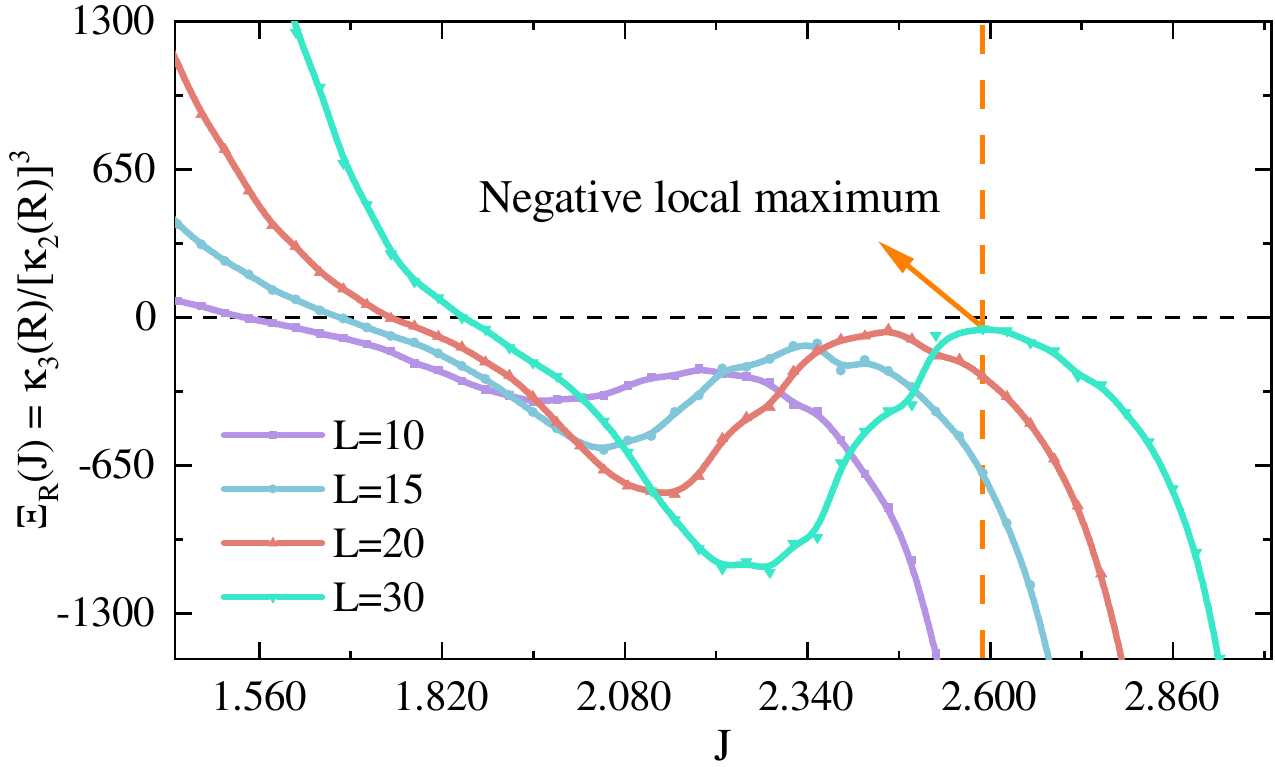}
\caption{Two-dimensional driven nonreciprocal Ising model at fixed $K_m=0.3$. $\Xi_R(J)$ is extracted from nonequilibrium steady-state time series of the synchronization observable $R(t)$ after discarding initial transients and performing block averaging. For $L=10,15,20,30$, $\Xi_R(J)$ develops a reproducible signed extremum and changes sign within a narrow $J$ window associated with the finite size onset of apparent synchronization and oscillatory transients. With increasing $L$, the feature gradually shifts and weakens, consistent with a crossover in two dimensions rather than a thermodynamic phase transition. This shows that the cumulant-ratio diagnostic remains fully operational without microcanonical input.}
  \label{fig:non}
\end{figure}

To probe higher-order fluctuation structure, we apply the cumulant-ratio diagnostic $\Xi_R(J)$ to the synchronization order parameter $R(t)$ as an operational tool. While the theoretical derivation of $\Xi(T)$ applies strictly to energy cumulants, this approach empirically captures precursor-like synchronization signals. All simulation parameters—including $K_m$, scan range $J$, total time steps, discarded transient period, and block averaging—follow Ref.~\cite{Avni2025NonreciprocalIsing} to ensure reproducibility. Smoothing in Fig.~\ref{fig:non} is used only for visualization; extrema are extracted from raw block-averaged time series.
As shown in Fig.~\ref{fig:non}, $\Xi_R(J)$ develops a reproducible signed extremum that changes sign within a narrow $J$ window associated with the finite size onset of apparent synchronization. With increasing $L$, the feature shifts, consistent with a crossover rather than a true thermodynamic phase transition. These results demonstrate that the cumulant-ratio diagnostic remains operational in nonequilibrium steady states, extending the fluctuation-based framework to physically meaningful nonequilibrium observables.

\emph{Conclusion---} We established a fluctuation-based canonical criterion for dependent and independent third-order transitions from the signed extrema of the cumulant ratio $\Xi(T)$. In the single-saddle regime, $N^2\Xi(T)\to s^{(3)}(e^\ast)$ provides a canonical--microcanonical bridge, linking fluctuation asymmetry to entropy-derivative structure. Physically, dependent signals mark precursor-like reorganizations on the disordered side of a major transition, whereas independent signals mark ordered-side restructuring; together they encode cooperative rearrangements, defect formation, and mesoscopic structural change beyond conventional low-order observables. Recent progress in direct microcanonical histogram analysis suggests that higher-order transitions form a resolvable hierarchy in finite systems~\cite{Bachmann2025Histogram}, and our results show that this hierarchy also admits a directly canonical formulation based only on measurable fluctuations. In equilibrium, $\Xi(T)$ complements microcanonical approaches without explicit density-of-states reconstruction, while in nonequilibrium steady states it remains operational when $g(E)$ and $S(E)$ are unavailable. Benchmarks on Ising, Potts, and driven nonreciprocal Ising models support an ensemble-bridging view of higher-order transitions in equilibrium and driven many-body systems.

\begin{acknowledgments}
This work was supported by China's National Natural Science Foundation Grant Nos.~12304257, 12322501, 12575035 and 12575033.
Y.T. acknowledges support from the Natural Science Foundation of Sichuan Province (Grant No.~2026NSFSCZY0124).
Computing resources were provided by the Interdisciplinary Intelligence Supercomputing Center of Beijing Normal University, Zhuhai.
\end{acknowledgments}

\bibliography{apssamp}

\end{document}


\title{Supplemental Material for ``Canonical Criterion for Third-Order Transitions''}

\author{Fangfang Wang$^{1,5}$}

\author{Wei Liu$^{2,1}$}
\email{weiliu@xust.edu.cn}
\thanks{Corresponding author.}

\author{Kai Qi$^{4}$}

\author{Zidong Cui$^{3}$}

\author{Ying Tang$^{3}$}
\email{jamestang23@gmail.com}
\thanks{Corresponding author.}

\author{Zengru Di$^{1,5}$}

\affiliation{%
$^1$Department of Systems Science, Faculty of Arts and Sciences, Beijing Normal University, Zhuhai 519087, China \\
$^2$College of Science, Xi'an University of Science and Technology, Xi'an 710600, China \\
$^3$Institute of Fundamental and Frontier Sciences, University of Electronic Science and Technology of China, Chengdu 611731, China \\
$^4$2020 X-Lab, Shanghai Institute of Microsystems and Information Technology,
Chinese Academy of Sciences, Shanghai 200050, China \\
$^5$School of Systems Science, Beijing Normal University, Beijing 100875, China
}

\maketitle

\tableofcontents

\section{Microcanonical Inflection-Point Analysis}
\label{sec:sm_mipa}

In this section, we briefly review the microcanonical inflection-point analysis (MIPA) and summarize the sign/extremum criteria used in the main text as the microcanonical reference for benchmarking the canonical diagnostic. 
Qi and Bachmann introduced MIPA by combining microcanonical thermodynamics with the principle of minimal sensitivity~\cite{Stevenson1981OPT,Stevenson1981RSA} to identify and classify first- and higher-order transitions in complex systems~\cite{Corti2023,Rocha2025MIPAParametric,wangff,liu2022pseudo,BelHadjAissa2021KT_XY}. 
The key idea is that, in finite systems, transition signatures need not appear as true thermodynamic singularities, but can still be identified from structured curvature changes in the microcanonical entropy and its derivative hierarchy~\cite{Gross2001MicrocanonicalThermodynamics,Junghans2006PeptideAggregation,Schnabel2011EntropyInflection}. 
Within an energy interval associated with a single phase, the microcanonical entropy and its derivatives vary smoothly; a transition is then signaled by a least-sensitive inflection point, namely an inflection point whose location changes minimally under small variations of the analysis window or numerical resolution~\cite{Qi2018MIPAClassification}.

The microcanonical entropy is defined as
\begin{equation}
S(E)=k_{\mathrm{B}}\ln g(E),
\end{equation}
where $k_{\mathrm{B}}$ is the Boltzmann constant and $g(E)$ is the density of states (DOS). 
To match the notation in the main text, we write the microcanonical inverse temperature as
\begin{equation}
\beta_\mu(E)\equiv \frac{\dd S(E)}{\dd E}.
\end{equation}
For brevity we also use $\beta(E)\equiv \beta_\mu(E)$ below. 
In the thermodynamically relevant energy range, $S(E)$ is a monotonically increasing function, and in the absence of transition signals its derivative hierarchy follows the characteristic monotonicity and curvature pattern discussed in Ref.~\cite{Qi2018MIPAClassification}. 
MIPA identifies transitions precisely through deviations from this regular pattern.

In MIPA, the transition location is therefore not defined by a global singularity criterion but by a local least-sensitivity condition in the derivative hierarchy. Operationally, the characteristic local extremum or inflection point is identified as the one whose position is least sensitive to small changes in the analysis window or in the numerical differentiation procedure. This is particularly useful in finite systems, where true thermodynamic singularities are rounded and transitions appear only as smooth but still clearly structured features in $S(E)$ and its derivatives.

If $S(E)$ exhibits a least-sensitive inflection point, $\beta(E)$ shows a positive local minimum, indicating an independent first-order transition. 
Similarly, if $\beta(E)$ exhibits a least-sensitive inflection point, the next derivative
\begin{equation}
\gamma(E)=\frac{\dd \beta(E)}{\dd E}=\frac{\dd^{2}S(E)}{\dd E^{2}}
\end{equation}
develops a characteristic extremum associated with an independent second-order transition, typically a negative local maximum. 
Proceeding further,
\begin{equation}
\delta(E)=\frac{\dd \gamma(E)}{\dd E}=\frac{\dd^{3}S(E)}{\dd E^{3}}
\end{equation}
can be used to detect third-order transitions. 
For an independent third-order transition, $\delta(E)$ develops a positive local minimum at the transition energy.

In addition to independent transitions, MIPA identifies dependent transitions, which occur only in coexistence with a lower-order independent transition and typically on the higher-energy side of the associated transition region. 
They indicate subtle precursor-like cooperative reorganization even in finite size systems, highlighting physical significance beyond the background monotonic derivative trend.
As emphasized in the original MIPA formulation, such dependent transitions are not generic consequences of the background monotonic trend of the derivative curves but additional least-sensitive features accompanying an independent transition. 
Because they occur at energies higher than those of the corresponding independent transition, they can be interpreted as precursor-like features in the disordered phase. 

The odd/even structure summarized in Table~\ref{tab:mipa_evenodd} reflects the alternating sign pattern in the derivative hierarchy of the microcanonical entropy. 
For odd-order transitions, classification follows local extrema of odd derivatives; for even-order, from even derivatives. Signs and extrema types must be interpreted together to ensure consistent classification across different derivative orders.
For odd-order transitions, the classification is read from local extrema of odd derivatives of $S(E)$, whereas for even-order transitions it is read from extrema of even derivatives. 
In practice, the sign and the extremum type must be interpreted together: the same sign can have different meanings at different derivative orders, while the odd/even grouping provides the consistent classification rule. 
For finite size systems, these criteria should be understood as a classification of structured local features rather than asymptotic singularities. 
Accordingly, whether a given transition is resolved in practice depends on system size, numerical resolution, and the quality of the underlying DOS estimate or derivative reconstruction.

\begin{table*}
\centering
\caption{Microcanonical MIPA sign criteria for independent and dependent transitions. The classification is evaluated at the transition energy $E_{\mathrm{tr}}$ and grouped by even and odd orders in the generalized inflection-point hierarchy. These criteria serve as the microcanonical reference for benchmarking the canonical diagnostic in the main text.}
\label{tab:mipa_evenodd}
\begin{tabular}{ccc}
\hline
Category & Even-order transitions ($2k$) & Odd-order transitions ($2k{-}1$) \\
\hline
Independent &
$\dd^{2k}S(E)/\dd E^{2k}<0$ (negative local maximum) &
$\dd^{2k-1}S(E)/\dd E^{2k-1}>0$ (positive local minimum) \\
Dependent &
$\dd^{2k}S(E)/\dd E^{2k}>0$ (positive local minimum) &
$\dd^{2k-1}S(E)/\dd E^{2k-1}<0$ (negative local maximum) \\
\hline
\end{tabular}
\end{table*}

In the main text, these microcanonical sign criteria provide the reference classification for benchmarking the canonical cumulant-ratio diagnostic. 
The microcanonical--canonical bridge derived in the next section shows how the corresponding canonical signatures inherit the same classification in the single-saddle regime.

\section{Microcanonical--Canonical Bridge}
\label{sec:sm_bridge}

In this section, we derive the asymptotic bridge between the microcanonical derivative hierarchy and the canonical cumulant-ratio diagnostic in the single-saddle regime. The goal is to show that the canonical ratio $\Xi(T)$ inherits the same third-order classification as the microcanonical quantity $s^{(3)}(e)$, thus providing the theoretical basis for the signed-extrema criterion used in the main text. Throughout this section, we set $k_{\mathrm B}=1$.

\subsection{Bridge between \texorpdfstring{$s^{(3)}(e^)$ and $\Xi(T)$}{s^(3)(e*) and Xi(T)}}

The microcanonical entropy per degree of freedom is
\begin{equation}
s(e)=\frac{S(E)}{N}=\frac{\ln g(E)}{N},
\end{equation}
with $E=Ne$. To distinguish the microcanonical derivative hierarchy from the canonical inverse temperature $\beta$, we define
\begin{equation}
\beta_\mu(e)=\frac{\dd s}{\dd e},\qquad
\gamma(e)=\frac{\dd^2 s}{\dd e^2},\qquad
\delta(e)=\frac{\dd^3 s}{\dd e^3}.
\end{equation}
On the physical concave branch, $\gamma(e)\le 0$. The third derivative $\delta(e)$ controls the leading asymmetry of the local entropy expansion around the saddle and therefore the asymmetry of the corresponding canonical energy fluctuations.

The canonical partition function can be written as
\begin{equation}
Z(\beta)=\sum_E g(E)e^{-\beta E}
\simeq
\int \dd e\, \exp\!\big[N\,\phi(e,\beta)\big],
\end{equation}
where
\begin{equation}
\phi(e,\beta)=s(e)-\beta e.
\end{equation}
In the single-saddle regime and for large $N$, the integral is dominated by a unique maximum at $e^\star(\beta)$ satisfying
\begin{equation}
\partial_e \phi(e^\star,\beta)=0
\quad\Longleftrightarrow\quad
\beta=\beta_\mu(e^\star)=s'(e^\star).
\label{eq:sm_saddle_condition}
\end{equation}
In this regime, the canonical ratio $\Xi(T)$ captures the asymmetry of energy fluctuations, reflecting cooperative rearrangements and providing a direct bridge to microcanonical derivatives $s^{(3)}(e^*)$.
Thus the canonical inverse temperature is matched to the microcanonical slope at the saddle-point energy density.

The canonical cumulants are generated from $K(\beta)=\ln Z(\beta)$:
\begin{equation}
\langle E\rangle=-\partial_\beta K,\qquad
\kappa_2=\partial_\beta^2 K,\qquad
\kappa_3=-\partial_\beta^3 K.
\end{equation}
In the single-saddle regime, one has $\langle E\rangle \approx N e^\star(\beta)$.

Differentiating Eq.~\eqref{eq:sm_saddle_condition} with respect to $\beta$ gives
\begin{equation}
1=\gamma(e^\star)\,\frac{\dd e^\star}{\dd\beta},
\end{equation}
and therefore
\begin{equation}
\frac{\dd e^\star}{\dd\beta}=\frac{1}{\gamma(e^\star)}.
\label{eq:sm_de_dbeta}
\end{equation}
Using $\langle E\rangle \approx N e^\star(\beta)$, the second cumulant becomes
\begin{equation}
\kappa_2
=
-\frac{\dd \langle E\rangle}{\dd\beta}
\approx
-N\,\frac{\dd e^\star}{\dd\beta}
\simeq
-\frac{N}{\gamma(e^\star)}.
\label{eq:sm_kappa2_bridge}
\end{equation}
Because $\gamma(e^\star)<0$ on the concave branch, $\kappa_2>0$ as required.

Differentiating once more, we obtain
\begin{equation}
\kappa_3
=
-\frac{\dd \kappa_2}{\dd\beta}
\approx
-\frac{\dd}{\dd\beta}\!\left(-\frac{N}{\gamma(e^\star)}\right).
\end{equation}
Applying the chain rule together with Eq.~\eqref{eq:sm_de_dbeta},
\begin{equation}
\kappa_3
\simeq
-\,N\,\frac{\dd}{\dd e}\!\left(\frac{1}{\gamma(e)}\right)_{e^\star}
\frac{\dd e^\star}{\dd\beta}.
\end{equation}
Since
\begin{equation}
\frac{\dd}{\dd e}\!\left(\frac{1}{\gamma(e)}\right)
=
-\frac{\gamma'(e)}{\gamma(e)^2}
=
-\frac{\delta(e)}{\gamma(e)^2},
\end{equation}
we obtain
\begin{equation}
\kappa_3
\simeq
-\,\frac{N\,\delta(e^\star)}{[\gamma(e^\star)]^3}.
\label{eq:sm_kappa3_bridge}
\end{equation}

Equation~\eqref{eq:sm_kappa3_bridge} is the microcanonical--canonical bridge for the third-order diagnostic. Since $\gamma(e^\star)<0$, the denominator $[\gamma(e^\star)]^3$ is negative, so $\kappa_3$ has the same sign as $\delta(e^\star)$. Hence
\begin{equation}
\delta(e_{\mathrm{ind}})>0 \;\Rightarrow\; \kappa_3(T_{\mathrm{ind}})>0,
\qquad
\delta(e_{\mathrm{dep}})<0 \;\Rightarrow\; \kappa_3(T_{\mathrm{dep}})<0.
\label{eq:sm_sign_rule}
\end{equation}
Because $\kappa_2>0$, the normalized ratio
\begin{equation}
\Xi(T)=\frac{\kappa_3(T)}{[\kappa_2(T)]^3}
\end{equation}
inherits the same sign structure. Therefore, in the single-saddle regime, the canonical signed-extrema rule is asymptotically equivalent to the microcanonical MIPA classification.

Combining Eqs.~\eqref{eq:sm_kappa2_bridge} and \eqref{eq:sm_kappa3_bridge} further gives
\begin{equation}
N^2\Xi(T)
=
N^2\frac{\kappa_3}{\kappa_2^3}
\simeq
\delta(e^\star)
=
s^{(3)}(e^\star),
\label{eq:sm_Xi_bridge}
\end{equation}
which is the asymptotic relation quoted in the main text and motivates the cubic normalization used in the definition of $\Xi(T)$.

Finally, using the exact canonical identity
\begin{equation}
\kappa_2(T)=T^2 C_V(T),
\end{equation}
together with $\kappa_3(T)=-\dd\kappa_2/\dd\beta$ and $\dd/\dd\beta=-(1/T^2)\,\dd/\dd T$, we obtain
\begin{equation}
\kappa_3(T)=T^4\frac{\dd C_V}{\dd T}+2T^3 C_V(T).
\label{eq:sm_kappa3_CV}
\end{equation}
Finite size deviations only slightly shift the extrema temperatures, without affecting the existence or qualitative identification of third-order transitions.
Equivalently,
\begin{equation}
\frac{\dd C_V}{\dd T}=\frac{\kappa_3}{T^4}-\frac{2C_V}{T}.
\end{equation}
This exact relation shows that the third cumulant encodes not only the magnitude of energy fluctuations through $C_V$, but also their temperature-dependent reorganization through $\dd C_V/\dd T$.

\subsection{Thermodynamic-limit Onsager reference temperatures}

As an application of the bridge, we evaluate $\Xi(T)$ for the thermodynamic-limit Onsager Ising solution \cite{Onsager1944} and extract the characteristic temperatures by the same local-fit protocol used throughout this work. We obtain
$T_{\mathrm{dep}}=2.567$,
$T_c=2.269$, and
$T_{\mathrm{ind}}=2.229$,
in agreement with the corresponding microcanonical benchmark values quoted in the main text. The resulting characteristic temperatures are summarized in Table~\ref{tab:ising_temps}.

\begin{table}[t]
\caption{\label{tab:ising_temps}
Transition temperatures of the two-dimensional Ising model from canonical and microcanonical analyses.}
\begin{tabular}{lccc}
\hline
 & $T_{\mathrm{dep}}$ & $T_c$ & $T_{\mathrm{ind}}$ \\
\hline
Microcanonical (MIPA) \cite{Qi2018MIPAClassification,Sitarachu2022IsingThirdOrder} & 2.567 & 2.269 & 2.229 \\
Canonical (this work) & 2.567 & 2.269 & 2.229 \\
\hline
\end{tabular}
\end{table}

These Onsager-based values provide the thermodynamic-limit reference temperatures used in Fig.~2 of the main text. 
By contrast, the geometry-sensitive observables in Fig.~2(b,c) are obtained from independent finite size Monte Carlo simulations using Wolff cluster updates, 
in order to collect detailed cluster statistics. 
Small offsets between the finite size simulation-based estimates and the thermodynamic-limit Onsager values are therefore expected, 
and they naturally reflect finite size rounding effects rather than indicating any inconsistency of the canonical criterion.
This comparison also highlights that the canonical criterion is not tied to a single numerical route: the same characteristic temperatures can be benchmarked against exact thermodynamic-limit input while their physical interpretation is visualized independently through finite-size mesoscopic observables.

\section{Geometry-sensitive observables and detailed analysis of Ising restructuring}
\label{sec:geometry_expanded}

In this section, we provide a more detailed discussion of the geometry-sensitive observables used to visualize mesoscopic structural changes associated with the canonical third-order transitions identified in the two-dimensional Ising benchmark~\cite{Sitarachu2022IsingThirdOrder,Liu2025PottsGeometry}. These observables are introduced purely as complementary probes and do not define independent transition criteria. They serve as quantitative indicators linking canonical extrema to mesoscopic cluster dynamics, while highlighting precursor-like cooperative reorganization on the disordered side of the transition region. The goal is to illustrate more explicitly the microscopic mechanisms underlying the canonical cumulant-ratio signature $\Xi(T)$, emphasizing the distinct structural reorganizations on the ordered and less-ordered sides of the critical region.

For each instantaneous lattice configuration, geometric spin clusters are constructed as connected components of nearest-neighbor sites in the same local state. The construction is purely geometric, implemented with periodic boundary conditions on the square lattice. Denoting the local state at site $i$ by $s_i$, where $s_i=\pm1$ for Ising-type binary fields, two nearest-neighbor sites belong to the same cluster if and only if
\begin{equation}
\delta_{s_i,s_j}=1, \qquad j\in \mathrm{nn}(i).
\end{equation}
Connected components are identified using standard labeling procedures. The physical interpretation of the clusters does not depend on the specific implementation.

For each cluster $C$, we record its area
\begin{equation}
A_C = \sum_{i\in C} 1,
\end{equation}
and if a configuration contains $N_{\mathrm{cl}}$ clusters, the cluster-averaged area is
\begin{equation}
A \equiv \frac{1}{N_{\mathrm{cl}}} \sum_C A_C.
\end{equation}
We also consider the number of isolated spins
\begin{equation}
n_1 = \sum_C \delta_{A_C,1},
\end{equation}
which serves as a sensitive measure of defects in the ordered phase.

In the Ising lattice, the independent transition at $T_{\rm ind}$ corresponds to a reorganization on the ordered side. 
Here, $A$ remains large, reflecting the stability of the dominant clusters, but its derivative $dA/dT$ exhibits a clear peak, indicating small-scale coalescence and rearrangement of domain boundaries. 
Concurrently, $n_1$ decreases rapidly, signaling that isolated spins (defects) are absorbed into larger domains as the ordered structure stabilizes. 
This microscopic behavior provides a visual and quantitative interpretation of the positive local minimum in $\Xi(T)$ observed at $T_{\rm ind}$.

By contrast, the dependent transition at $T_{\rm dep}$ occurs on the less-ordered side of the major transition. 
In this region, $n_1$ shows a transient increase with temperature, reflecting the emergence of local defects and precursor fluctuations in smaller clusters. 
Simultaneously, the derivative $dA/dT$
may show a modest positive or negative feature, indicating that small clusters are transiently forming and dissolving. 
These structural fluctuations are consistent with the negative local maximum of $\Xi(T)$ at $T_{\rm dep}$ and provide a microscopic picture of precursor-like cooperative reorganization in the disordered phase.

The combination of $A$, $dA/dT$, and $n_1$ across the full temperature range reveals a clear asymmetry between the ordered and less-ordered sides of the critical region. On the ordered side corresponding to the independent transition, large clusters dominate, $dA/dT$ exhibits a pronounced peak, and isolated spins are progressively absorbed into larger clusters, corresponding to the positive minimum of $\Xi(T)$. On the less-ordered side associated with the dependent transition, small clusters fluctuate more strongly, $n_1$ rises transiently, and $dA/dT$ shows relatively minor features, corresponding to the negative maximum of $\Xi(T)$. These geometry-sensitive observables do not replace $\Xi(T)$, but instead provide an intuitive and quantitative mapping from canonical fluctuation extrema to the underlying microscopic cluster dynamics. They allow one to visualize more clearly how dependent and independent third-order transitions manifest as distinct reorganizations of mesoscopic structure in the Ising lattice, thus bridging the gap between statistical diagnostics and underlying microscopic physics. Because these observables are model-dependent, they are not intended as universal transition criteria. Their role is instead interpretive: they make visible the mesoscopic restructuring that the universal canonical diagnostic $\Xi(T)$ already detects.

\section{Potts model: numerical methods}
\label{sec:potts-methods}

In this section, we describe the numerical procedures used to evaluate the canonical diagnostic in finite size Potts models~\cite{Potts1952}, including DOS estimation, reweighting, numerical conditioning, and the extraction of characteristic temperatures. To further demonstrate the versatility of the canonical diagnostic, we evaluate $\Xi(T)$ starting from the density of states $g(E)$. Although $\Xi(T)$ is defined in terms of canonical energy cumulants, it can be computed either directly from canonical sampling of energy fluctuations or from a microcanonical input $g(E)$ via standard reweighting. Here we adopt the latter route and estimate $g(E)$ using a replica Wang--Landau~\cite{Wang2001MultipleRangeRW,Vogel2013Hierarchical,Vogel2014ReplicaExchange} scheme with energy-window decomposition.

The total energy range $[E_{\min},E_{\max}]$ is divided into $n_w$ windows with $10\%$ overlap at both ends. Typical choices are $n_w\in[12,24]$ for $L=30$ to $56$. Each window hosts one Wang--Landau walker. Neighboring replicas attempt window exchanges every $10^3$ sweeps using a Metropolis acceptance based on the current DOS estimates. We initialize the modification factor at $f_0=e^1$ and update it according to $f\leftarrow \sqrt{f}$ between stages until $\ln f<10^{-8}$. Flatness is monitored using the relative histogram criterion
\[
\min_E H(E)/\langle H\rangle>0.8,
\]
supplemented by the integrated autocorrelation time $\tau_{\mathrm{int}}$ to avoid false-flat plateaus. In the overlap regions, adjacent estimates of $\ln g(E)$ are stitched by weighted least squares with weights proportional to $H(E)$. The additive constant of $\ln g(E)$ is fixed using the exact partition function at $\beta=0$, namely $Z(0)=q^N$ with $N=L^2$. For each system size we perform $N_{\mathrm{seed}}=10$ independent runs. Within each Wang--Landau stage, the sweep length is chosen to be at least $50\,\tau_{\mathrm{int}}$. Uncertainties are propagated to derived quantities, including $\Xi(T)$, by block-jackknife with block size no smaller than $2\,\tau_{\mathrm{int}}$.

Given a stitched DOS $g(E)$, canonical moments at inverse temperature $\beta$ are obtained by reweighting:
\begin{equation}
\langle E^n\rangle_{\beta}=
\frac{\sum_{E} E^n\,g(E)\,e^{-\beta E}}{\sum_{E} g(E)\,e^{-\beta E}}.
\end{equation}
The second and third energy cumulants are then
\begin{align}
\kappa_2(\beta)&=\langle E^2\rangle_\beta-\langle E\rangle_\beta^2,\\
\kappa_3(\beta)&=\langle (E-\langle E\rangle_\beta)^3\rangle_\beta
=\langle E^3\rangle_\beta-3\langle E^2\rangle_\beta\langle E\rangle_\beta+2\langle E\rangle_\beta^3,
\end{align}
and the canonical diagnostic is
\begin{equation}
\Xi(T)=\frac{\kappa_3(T)}{[\kappa_2(T)]^3},
\qquad
\beta=\frac{1}{T}.
\end{equation}

Because $\kappa_3$ is more sensitive to the far distribution tails than $\kappa_2$, the reliable temperature range for $\Xi(T)$ is generally narrower than that for lower-order observables. In practice, we report and fit $\Xi(T)$ only within the interval where the reweighting weights remain numerically and statistically well conditioned and the dominant support of the canonical energy distribution is covered by well-sampled DOS regions. Operationally, we monitor the effective overlap between the reweighting kernel $g(E)e^{-\beta E}$ and the stitched DOS support obtained from the Wang--Landau windows. The reliable temperature interval for $\Xi(T)$ is therefore restricted to regions where the reweighted distribution is well sampled and sufficiently overlaps the dominant DOS support. Temperatures at which the reweighted distribution becomes dominated by poorly sampled edge bins are excluded from the extremum and crossing analysis. This conditioning step is particularly important in finite-size first-order regimes, where coexistence-induced asymmetry further enhances the tail sensitivity of $\kappa_3$. In such regimes, the reweighted energy distribution may broaden substantially or even develop incipient two-peak structure, so small inaccuracies near the sampled DOS boundaries can propagate disproportionately into $\kappa_3$ and hence into $\Xi(T)$. Conservative temperature-window selection is therefore essential for stable extremum extraction and size-crossing analysis.

\section{Potts-model finite size benchmarks and comparison with MIPA}
\label{sec:potts-results}

In this section, we benchmark the canonical diagnostic in finite size square-lattice Potts models and compare the extracted characteristic temperatures with the available microcanonical MIPA references. Starting from the stitched DOS, we compute canonical cumulants by reweighting and evaluate the same diagnostic $\Xi(T)$ used in the main text. This provides an equilibrium benchmark of the canonical fluctuation framework through a DOS-based route.

\begin{figure*}[t]
  \centering
  \includegraphics[width=0.98\textwidth]{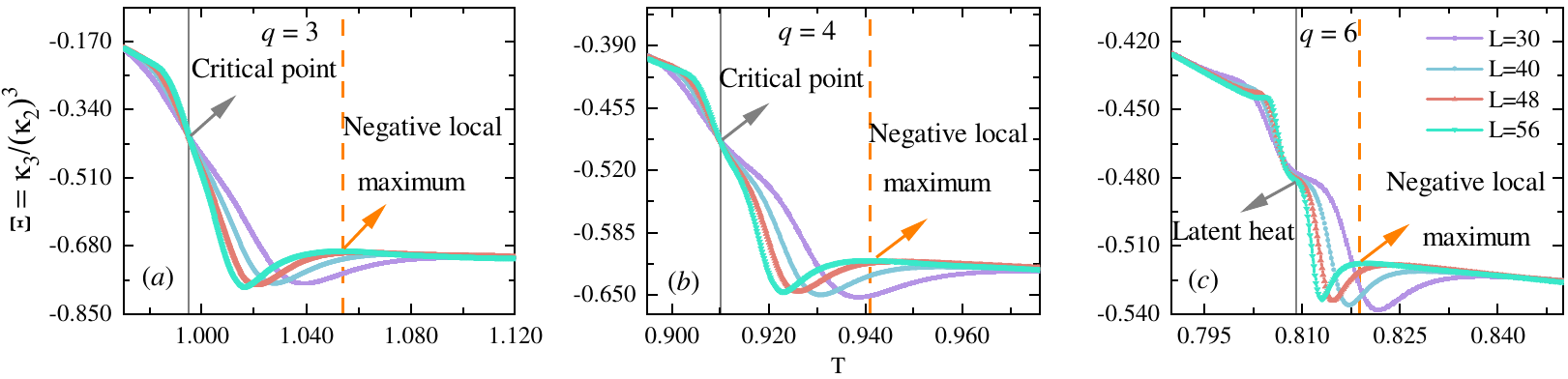}
  \caption{Third-order transitions in square-lattice Potts models. From the stitched DOS we obtain canonical cumulants by reweighting and evaluate $\Xi(T)=\kappa_3/(\kappa_2)^3$ for $q=3,4,6$. In (a)--(c), the negative local maximum identifies the dependent transition at $T_{\rm dep}=1.054$ ($q=3$), $0.941$ ($q=4$), and $0.820$ ($q=6$). Here we report the same $L = 56$ estimates used for the dashed/dotted markers in the main-text figure. Solid lines mark $T_c/J=1/\ln(1+\sqrt{q})$.}
  \label{fig:potts}
\end{figure*}

\begin{table}[t]
  \centering
  \caption{Square-lattice Potts critical temperatures from microcanonical, exact, and canonical analyses. Canonical values are finite size estimates obtained from local fits to size crossings of $\Xi(T)$.}
  \begin{tabular}{cccc}
  \hline
    $q$ & $T_c$ (micro) \cite{wangff} & $T_c$ (exact) \cite{Potts1952} & $T_c$ (can) \\
    \hline
    3 & 1.000 & 0.995 & 0.999 \\
    4 & 0.914 & 0.910 & 0.913 \\
    6 & 0.809 & 0.808 & 0.809 \\
    8 & 0.744 & 0.745 & 0.746 \\
    \hline
  \end{tabular}
  \label{tab:potts-tc}
\end{table}

\begin{table}[t]
  \centering
  \caption{Square-lattice Potts third-order transitions (NF: no corresponding transition feature resolved within the analyzed finite size window in the specified ensemble). Canonical values are finite size estimates extracted from local fits to signed extrema of $\Xi(T)$; for $q=8$, the listed canonical values correspond to the same $L=56$ estimates used for the markers in the main-text figure.}
  \begin{tabular}{ccccc}
  \hline
    $q$ & $T_{\rm dep}$ (micro) \cite{wangff} & $T_{\rm dep}$ (can) & $T_{\rm ind}$ (micro) & $T_{\rm ind}$ (can) \\
    \hline
    3 & 1.050 & 1.054 & NF & NF \\
    4 & 0.935 & 0.941 & NF & NF \\
    6 & 0.811 & 0.820 & 0.811 & NF \\
    8 & NF    & 0.748 & 0.742 & 0.742 \\
    \hline
  \end{tabular}
  \label{tab:potts-third}
\end{table}

Supplementary Figure~\ref{fig:potts} shows that $\Xi(T)$ retains clear finite size crossings and signed extrema across both the continuous ($q\le 4$) and first-order ($q>4$) regimes. For $q=6$, the curves develop a pronounced step-like profile, consistent with latent heat and thus characteristic of first-order behavior. The size crossing of $\Xi(T)$ provides an accurate estimator of $T_c$, in close agreement with the exact result $T_c/J=1/\ln(1+\sqrt{q})$ (Table~\ref{tab:potts-tc}), while the signed extrema yield the characteristic third-order temperatures listed in Table~\ref{tab:potts-third}. Where the corresponding microcanonical features are resolved, the canonical temperatures are consistent with the available MIPA benchmarks.
NF entries indicate unresolved features due to finite size limitations, not the absence of precursor-like structure. 

These results confirm $\Xi(T)$ provides a stable measure of third-order transitions across system sizes and ensemble definitions.
These observations are consistent with the MIPA benchmarks: the third-order transitions observed in finite size $q=6$ and $q=8$ Potts models, as well as in finite size Ising systems~\cite{Sitarachu2020ThirdOrderIsing}, persist across lattice sizes. Similar precursor-like signals have been reported in protein-folding models~\cite{Qi2019FlexiblePolymers,DiCairano2022Topological}, in spin models~\cite{liu2022pseudo,wangff,liu2025PLA,shi2026pseudo,Liu2025PottsGeometry,BelHadjAissa2021KT_XY}, and in majority-vote dynamics~\cite{Liu2024MV}. Moreover, the exact density-of-states solution confirms that both independent and dependent third-order transitions remain robust in the thermodynamic limit~\cite{Sitarachu2022IsingThirdOrder}, indicating that these signals are genuine physical features rather than finite size artifacts.

The crossing-based estimate is particularly robust because it relies on the size-dependent reshaping of energy fluctuations rather than on resolving a singularity directly, and therefore remains effective even under substantial finite size. In this sense, $\Xi(T)$ complements standard thermodynamic diagnostics by encoding higher-cumulant information that is sensitive to coexistence-induced asymmetry of the energy distribution in first-order regimes. Within the MIPA classification, dependent transitions occur only in coexistence with a lower-order independent transition and therefore naturally appear on the higher-temperature (higher-energy) side as precursor features of the associated major transition. Accordingly, entries labeled ``NF'' in Table~\ref{tab:potts-third} indicate that the corresponding transition feature is not resolved within the analyzed finite size window, even though precursor-like structure may still remain detectable.

Taken together, these Potts benchmarks show that $\Xi(T)$ can be evaluated not only from canonical fluctuation data but also from a DOS input via reweighting, while preserving the canonical fluctuation picture developed in the main text. The system sizes reported here are chosen to resolve the finite size crossing and signed-extremum structure of $\Xi(T)$ with controlled DOS-reweighting statistics over dense temperature scans. Extending to substantially larger $L$ is feasible in principle, but would require a disproportionate increase in Wang--Landau sampling effort together with tighter control of tail-conditioned reweighting for $\kappa_3$, and is therefore beyond the scope of the present finite size benchmark.

\section{General-order canonical diagnostic and operational definition}
\label{sec:general}

In this section, we outline a general-order extension of the canonical diagnostic and summarize the corresponding operational sign criteria beyond the third-order case. The approach reconstructs the microcanonical curvature hierarchy from canonical energy cumulants in the single-saddle regime, where the Laplace transform of $g(E)$ is dominated by a unique saddle energy density $e^\ast(T)$. In the thermodynamic limit, the saddle is determined by $s'(e^\ast)=\beta=1/T$ with $s(e)=S(E)/N$ and $E=Ne$, and we denote $s^{(k)}\equiv \partial_e^k s(e^\ast)$.

A standard saddle expansion gives the leading-order relation between energy cumulants and derivatives of $s(e^\ast)$:
\begin{align}
\kappa_2 &\simeq -\frac{N}{s^{(2)}},\\
\kappa_3 &\simeq -N\,\frac{s^{(3)}}{[s^{(2)}]^3},\\
\kappa_4 &\simeq N\left(\frac{3[s^{(3)}]^2}{[s^{(2)}]^5}-\frac{s^{(4)}}{[s^{(2)}]^4}\right),\ \dots
\end{align}

To extend the canonical diagnostic to general order $n\ge 3$, we define the operator
\begin{equation}
D\equiv -\frac{N}{\kappa_2(\beta)}\frac{\mathrm d}{\mathrm d\beta},
\end{equation}
and reconstruct the derivative hierarchy from canonical cumulants as
\begin{equation}
\tilde{s}_d^{(2)}(\beta)\equiv -\frac{N}{\kappa_2(\beta)},\qquad
\tilde{s}_d^{(n)}(\beta)\equiv D^{\,n-2}\tilde{s}_d^{(2)}(\beta)\ \ (n\ge 3),
\end{equation}
so that in the single-saddle thermodynamic limit, $\tilde{s}_d^{(n)}(\beta)\to s^{(n)}(e^\ast(\beta))$. The normalized $n$th-order diagnostic is then
\begin{equation}
\Sigma_n(T)\equiv \frac{1}{N^{\,n-1}}\tilde{s}_d^{(n)}(\beta)\big|_{\beta=1/T},
\end{equation}
which satisfies
\begin{equation}
N^{\,n-1}\Sigma_n(T)\longrightarrow s^{(n)}(e^\ast),
\end{equation}
and reduces to the third-order canonical--microcanonical bridge for $n=3$, i.e., $\Sigma_3(T)=\Xi(T)$.

The first few explicit forms constructed solely from energy cumulants are
\begin{align}
\Sigma_3 &= \frac{\kappa_3}{\kappa_2^3}\equiv \Xi,\\
\Sigma_4 &= \frac{\kappa_2\kappa_4-3\kappa_3^2}{\kappa_2^5},\\
\Sigma_5 &= \frac{\kappa_2^2\kappa_5-10\kappa_2\kappa_3\kappa_4+15\kappa_3^3}{\kappa_2^7}.
\end{align}
Because higher-order cumulants are increasingly sensitive to noise and discretization, in practice we focus on $n=3$ for the main text.
From a practical perspective, the third-order case is the lowest order at which fluctuation asymmetry enters explicitly while still remaining numerically stable in finite-size data. For this reason, the higher-order extension is best viewed as a systematic completion of the framework and as a guide for future large-scale studies, rather than as a replacement for the third-order diagnostic used here. Moreover, this emphasis on the third-order case is not only numerical but also physical. As shown in the main text, the third-order diagnostic is the lowest-order canonical observable that already resolves the asymmetry of fluctuation reorganization around a major transition, distinguishing precursor-like disordered-side structure from ordered-side restructuring. The same sign-extremum criteria apply consistently across system sizes, ensuring operational identification of independent and dependent transitions beyond third order.
For larger $n$, stable extraction of $\Sigma_n(T)$ requires stronger smoothing, denser scans, or larger ensembles, particularly for nonequilibrium time-series data.

In the single-saddle limit, $N^{\,n-1}\Sigma_n(T)\to s^{(n)}(e^\ast)$, so the same sign-extremum criteria as in the generalized inflection-point analysis continue to apply consistently across system sizes. Operationally, for odd order ($n=2k-1$), an independent transition is identified by a positive local minimum and a dependent transition by a negative local maximum. For even order ($n=2k$), an independent transition is identified by a negative local maximum and a dependent transition by a positive local minimum. Using the relations $\kappa_2(T)=T^2 C_V(T)$ and $\kappa_3(T)=-\mathrm d\kappa_2/\mathrm d\beta$, with $\mathrm d/\mathrm d\beta=-(1/T^2)\mathrm d/\mathrm dT$, higher-order $\kappa_n$ can be obtained recursively. Consequently, $\Sigma_n(T)$ can be constructed directly from $C_V(T)$ and its temperature derivatives up to order $n-2$, without requiring any explicit order parameter or free-energy differences.

Taken together, the Supplemental Material provides the microcanonical reference framework, the single-saddle canonical--microcanonical bridge, the numerical protocols underlying the Ising and Potts benchmarks, and the general-order extension of the cumulant-based construction. These results clarify both the theoretical basis and the practical scope of the canonical criterion introduced in the main text. This equilibrium DOS-based route also clarifies an important conceptual point: although the input may be microcanonical, the extracted observable remains purely canonical because it is defined entirely through canonical cumulants. This is precisely what allows the same signed-extremum criterion to be transferred in a natural way to nonequilibrium steady-state time series, where no explicit density-of-states reconstruction is available. In practice, the third-order diagnostic $\Xi(T)$ offers the most robust overall balance between physical interpretability, numerical stability, and broad applicability to both equilibrium and nonequilibrium settings.

To provide a complete reference for readers, Supplementary Table~\ref{tab:sym} summarizes all symbols used throughout. 
This table serves as a convenient guide for cross-referencing definitions and first appearances of each symbol.

\begin{table}[t]
\centering
\caption{Glossary of symbols used in the canonical framework for third-order transitions.}
\begin{tabular}{ll}
\hline
Notation & Description / Physical meaning \\
\hline
$E$ & Total energy of the system (Hamiltonian) \\
$e$ & Energy density, $e = E/N$ \\
$N$ & Number of degrees of freedom \\
$S(E)$ & Microcanonical entropy, $S(E) = k_B \ln g(E)$ \\
$s(e)$ & Entropy per degree of freedom, $s(e) = S(E)/N$ \\
$\beta$ & Canonical inverse temperature, $\beta = 1/(k_B T)$ \\
$\beta_\mu(E)$ & Microcanonical inverse temperature, $\beta_\mu(E) = dS/dE$ \\
$\kappa_n$ & $n$th energy cumulant, $\kappa_n = (-1)^n \partial_\beta^n K(\beta)$ \\
$U$ & Average energy, $\langle E \rangle$ \\
$\Xi(T)$ & Third-order canonical diagnostic, $\Xi(T) = \kappa_3 / (\kappa_2)^3$ \\
$\Sigma_n(T)$ & General-order canonical diagnostic \\
$s^{(n)}(e)$ & $n$th derivative of the microcanonical entropy \\
$\gamma(e)$ & Second derivative of microcanonical entropy, $d^2 s / de^2$, controlling concavity \\
$\delta(e)$ & Third derivative of microcanonical entropy, $d^3 s / de^3$, controlling distribution asymmetry \\
$g(E)$ & Density of states (DOS) \\
$P(E)$ & Canonical energy distribution, $P(E) = g(E)e^{-\beta E}/Z(\beta)$ \\
$n_1$ & Number of isolated spins in lattice configurations \\
$A_C$ & Area of cluster $C$ \\
$N_{\rm cl}$ & Number of clusters in a configuration \\
$A$ & Average cluster area, $A = \sum_C A_C / N_{\rm cl}$ \\
$dA/dT$ & Temperature derivative of the average cluster area \\
$\sigma_i$ & Potts spin at site $i$, $\sigma_i \in \{0,1,\dots,q-1\}$ \\
$\sigma_i^{A}, \sigma_i^{B}$ & Nonreciprocal Ising spins at site $i$, taking values $\pm 1$ \\
$M_A$ & Magnetization of species $A$, $M_A = \frac{1}{N}\sum_i \sigma_i^A$ \\
$M_B$ & Magnetization of species $B$, $M_B = \frac{1}{N}\sum_i \sigma_i^B$ \\
$R(t)$ & Synchronization order parameter in driven Ising, $R(t) = \sqrt{M_A^2 + M_B^2}/\sqrt{2}$ \\
$K_m$ & Nonreciprocity strength in driven Ising \\
\hline
\end{tabular}
\label{tab:sym}
\end{table}

\bibliography{apssamp.bib}